# Robust exceptional point of arbitrary order in coupled spinning cylinders

Hongkang Shi[1,5], Zheng Yang[2,5], Chengzhi Zhang[2], Yuqiong Cheng[2], Yuntian Chen[1,3,6], and Shubo Wang[2,4,*]

[1]School of Optical and Electronic Information, Huazhong University of Science and Technology, Wuhan 430074, China
[2]Department of Physics, City University of Hong Kong, Tat Chee Avenue, Kowloon, Hong Kong, China
[3]Wuhan National Laboratory of Optoelectronics, Huazhong University of Science and Technology, Wuhan 430074, China
[4]City University of Hong Kong Shenzhen Research Institute, Shenzhen, Guangdong 518057, China

[5]These authors contributed equally
[6]yuntian@hust.edu.cn
*shubwang@cityu.edu.hk

**Abstract**

Exceptional points (EPs), i.e., non-Hermitian degeneracies at which eigenvalues and eigenvectors coalesce, can be realized by tuning the gain/loss contrast of different modes in non-Hermitian systems or by engineering the asymmetric coupling of modes. Here we demonstrate a mechanism that can achieve EPs of arbitrary order by employing the non-reciprocal coupling of spinning cylinders sitting on a dielectric waveguide. The spinning motion breaks the time-reversal symmetry and removes the degeneracy of opposite chiral modes of the cylinders. Under the excitation of a linearly polarized plane wave, the chiral mode of one cylinder can unidirectionally couple to the same mode of the other cylinder via the spin-orbit interaction associated with the evanescent wave of the waveguide. The structure can give rise to arbitrary-order EPs that are robust against spin-flipping perturbations, in contrast to conventional systems relying on spin-selective excitations. In addition, we show that higher-order EPs in the proposed system are accompanied by enhanced optical isolation, which may find applications in designing novel optical isolators, nonreciprocal optical devices, and topological photonics.

## 1. Introduction

While a Hermitian matrix must have real eigenvalues and orthogonal eigenvectors, the eigenvalues of a non-Hermitian matrix are complex in general. A special case is the non-Hermitian systems with parity-time (PT) symmetry. By tuning the complex potentials (e.g., gain and loss) of these systems, two or more eigenvalues can coalesce at one point in the parameter space. This point of degeneracy marks a spontaneous phase transition from real eigenvalues to complex and is called exceptional point (EP) [1–4]. Exceptional points are generic properties of the non-Hermitian Hamiltonians and can also emerge in purely lossy/gain systems without PT symmetry [5]. Numerous unexpected phenomena have been found near the EPs, and they have generated various applications such as optical isolation [6,7], optical switches [8–12], and unconventional lasers [5,13,14], etc. Recently, higher-order EPs have attracted considerable attention [15,16]. These EPs are characterized by more complicated topological energy surfaces [17,18] and ultrasensitive response to external perturbations [19,20]. Nevertheless, it remains challenging to realize higher-order EPs in general non-Hermitian systems with a large parameter space.

One approach to realize higher-order EPs is by employing the unidirectional coupling between multiple resonators. The mechanism can be understood by considering a simple two-level system described by the non-Hermitian Hamiltonian $H = \begin{bmatrix} a & c \\ d & b \end{bmatrix}$. An EP occurs at $(a-b)^2 + 4cd = 0$, which can be realized if $a$ and $b$ have equal real parts but different imaginary parts, and $c = d^*$. This scenario corresponds to conventional EPs induced by asymmetric loss/gain [21,22]. On the other hand, EPs can also appear when $a = b$ and either $c = 0$ or $d = 0$, which corresponds to two modes with unidirectional coupling [23,24]. Unidirectional coupling can be achieved by tuning the interferences of several perturbative scatterers [23,25]. It can also be realized by employing the spin-orbit interaction associated with evanescent waves of waveguide modes, where the propagating direction of waveguide modes is locked to the transverse spin of evanescent waves [26–30]. This method has successfully been applied to achieve arbitrary-order EPs in coupled dipole resonators [24], and it has a remarkable advantage of easy implementation without the need of tuning any parameters. However, this method relies on selective excitation of one chiral mode of the resonators. Since the system supports a pair of EPs due to time-reversal symmetry, subject to strong perturbations, the spin of the evanescent wave can be flipped, and the two EPs can couple with each other. This can affect the desired non-Hermitian properties of the systems.

In this paper, we demonstrate a mechanism that can realize EPs of arbitrary order by using two-dimensional (2D) spinning cylindrical resonators. The spinning motion breaks the time-reversal symmetry and induces a synthetic gauge field that splits a pair of frequency-degenerate chiral modes. At the frequency of one chiral mode, the coupling between the cylinders is unidirectional due to the spin-orbit interaction of the waveguide mode. The EP achieved in our system can be of arbitrary order depending on the number of spinning cylinders. Importantly, the broken time-reversal symmetry lifts the degeneracy of the chiral modes and thus avoids the effects of spin flipping induced by perturbations. Using full-wave numerical simulations and coupled-mode theory (CMT), we studied the modal properties of the spinning cylinders and the transmission properties of the waveguide at the EPs. Strong asymmetric response of the cylinders and the waveguide transmission were observed. We found that the order of the EPs is directly related to the optical isolation ratio in the structure.

The paper is organized as follows. In Section 2, we provide the formulations describing the nonreciprocal properties of spinning cylinders and the CMT used to understand the mode coupling between spinning cylinders. In Section 3, we present the numerical and analytical results with discussions. We then draw the conclusion in Section 4.

## 2. Formulations of the coupled spinning cylinders

*2.1 Effective constitutive relations and Sagnac frequency splitting of a single rotating cylinder*

We first consider a 2D dielectric cylinder with its axis aligned in $z$ direction and spinning with angular velocity $\mathbf{\Omega} = \Omega \hat{z}$. The electromagnetic properties of the cylinder can be described by the Minkowski constitutive relations [31]:

$$\mathbf{D} + \mathbf{v} \times \frac{\mathbf{H}}{c^2} = \boldsymbol{\varepsilon} \cdot (\mathbf{E} + \mathbf{v} \times \mathbf{B}),$$
$$\mathbf{B} + \mathbf{E} \times \frac{\mathbf{v}}{c^2} = \boldsymbol{\mu} \cdot (\mathbf{H} + \mathbf{D} \times \mathbf{v}), \tag{1}$$

where $\mathbf{E}$, $\mathbf{D}$, $\mathbf{B}$, $\mathbf{H}$ are the electromagnetic fields, $c$ is the speed of light in vacuum, $\mathbf{v}$ is the linear velocity of the cylinder, $\boldsymbol{\varepsilon}$ and $\boldsymbol{\mu}$ are the effective permittivity and permeability tensors, respectively. Equation (1) indicates that the spinning motion of the cylinder induces the coupling between electric and magnetic fields and turns the medium into an effective inhomogeneous and bi-anisotropic medium. The above constitutive relations can be rearranged as $\begin{bmatrix} \mathbf{D} \\ \mathbf{B} \end{bmatrix} = \begin{bmatrix} \boldsymbol{\varepsilon}' & \boldsymbol{\chi}_{em} \\ \boldsymbol{\chi}_{me} & \boldsymbol{\mu}' \end{bmatrix} \begin{bmatrix} \mathbf{E} \\ \mathbf{H} \end{bmatrix}$, where $\boldsymbol{\chi}_{em}$ and $\boldsymbol{\chi}_{me}$ characterize the bi-anisotropic properties of the medium. Different from the reciprocal medium with $\boldsymbol{\varepsilon}' = \boldsymbol{\varepsilon}'^T$, $\boldsymbol{\mu}' = \boldsymbol{\mu}'^T$ and $\boldsymbol{\chi}_{em} = (\boldsymbol{\chi}_{me}^*)^T = -\boldsymbol{\chi}_{me}^T$, we have $\boldsymbol{\varepsilon}' = \boldsymbol{\varepsilon}'^T$, $\boldsymbol{\mu}' = \boldsymbol{\mu}'^T$ and $\boldsymbol{\chi}_{em} = (\boldsymbol{\chi}_{me}^*)^T = \boldsymbol{\chi}_{me}^T$ for this spinning cylinder. Thus, the cylinder breaks time-reversal symmetry and is nonreciprocal [30,32]. Under TE polarization, the eigenmodes of the cylinder take the form of $E_z = E_z(r)\exp(\pm im\theta)$, where $m$ denotes the azimuthal quantum number of the chiral modes and $\theta$ is the azimuthal angle in the cylindrical coordinate system. The modes with $+m$ and $-m$ rotate in clockwise (CW) and counter-clockwise (CCW) directions, respectively. They carry spin angular momentum in $z$ direction and thus can couple with the transverse spin of evanescent waves in waveguide modes [30]. Applying the above constitutive relations to TE-polarized Helmholtz equation, in the limit of $\Omega R \ll c$, we can obtain the Sagnac frequency splitting between a pair of chiral modes (i.e. CW and CCW modes) with $\pm m$ [32]:

$$\Delta \omega = \frac{2m(\varepsilon_r \mu_r - 1)\Omega}{\varepsilon_r \mu_r}. \tag{2}$$

Here $\varepsilon_r$ and $\mu_r$ are the relative permittivity and relative permeability of the stationary cylinder, respectively. It is obvious that the frequency splitting is proportional to the azimuthal quantum number $m$ and angular velocity $\Omega$. For a stationary cylinder ($\Omega = 0$), we have $\Delta \omega = 0$ and the two chiral modes with $\pm m$ are degenerate. This degeneracy disappears for non-zero angular velocity Ω. The frequency splitting can be understood as resulting from the hybridization of the CW and CCW modes (in the stationary cylinder) induced by mirror symmetry breaking under spinning of the cylinder. It can also be understood as a result of the synthetic gauge field induced by the cylinder spinning [30]. We note that the frequency splitting is of critical importance to the realization of robust EPs in our systems, as it can suppress the direct coupling between CW and CCW modes due to spin flipping in presence of perturbations. We will elaborate on this point next.

*2.2 Coupled mode theory for two spinning cylinders sitting on a waveguide*

We now consider the coupling of the two spinning dielectric cylinders sitting above a dielectric slab waveguide and separated by a distance $d$, as shown in Fig. 1(a). The structure is excited by a plane wave propagating in $-y$ direction and linearly polarized in $z$ direction. The rate equations for the two coupled cylinders can be expressed as [24,33,34]:

$$\begin{aligned}
\frac{da_1}{dt} &= -i\omega_0 a_1 - \frac{\gamma_1 + \gamma_c}{2} a_1 - i\kappa_{12} a_2 - \sqrt{\gamma_c} a_{\text{in}}, \\
\frac{da_2}{dt} &= -i\omega_0 a_2 - \frac{\gamma_2 + \gamma_c}{2} a_1 - i\kappa_{21} a_1 - \sqrt{\gamma_c} a_{\text{in}},
\end{aligned} \quad (3)$$

where $\omega_0$ is the resonance frequency of the isolated cylinders; $\gamma_{1,2}$ and $\gamma_c$ are the material loss and radiation loss of the cylinders, respectively; $\kappa_{21}$ ($\kappa_{12}$) represents the coupling from cylinder 1 (2) to cylinder 2 (1); $a_1$, $a_2$ are the mode fields of cylinder 1 and cylinder 2, respectively; $a_{\text{in}}$ is the field of the incident wave. The above equation can be re-written as:

$$\frac{d\Lambda}{dt} = -iH\Lambda - \sqrt{\gamma_c}\Lambda_{\text{in}}, \quad (4)$$

where

$$H = \begin{bmatrix} \omega_0 - \frac{i}{2}(\gamma_1 + \gamma_c) & \kappa_{12} \\ \kappa_{21} & \omega_0 - \frac{i}{2}(\gamma_2 + \gamma_c) \end{bmatrix},$$

$$\Lambda = \begin{bmatrix} a_1 \\ a_2 \end{bmatrix}, \Lambda_{\text{in}} = \begin{bmatrix} a_{\text{in}} \\ a_{\text{in}} \end{bmatrix}. \quad (5)$$

For the chiral CW (CCW) mode of the spinning cylinders, its eigen magnetic field carries spin angular momentum in the $-z$ ($+z$) direction [30]. This spin matches the transverse spin of the evanescent wave of the waveguide mode and is locked to the propagating direction of the waveguide mode [28]. Consequently, the coupling between the cylinders is unidirectional. Thus, we have $\kappa_{12} = 0$ and $\kappa_{21} \neq 0$, and an EP appears at $\gamma_1 = \gamma_2 = \gamma$, corresponding to identical cylinders. Assuming the time-harmonic mode field $a_i = A_i \exp(-i\omega t)$ and incident field $a_{in} = A_{in} \exp(-i\omega t)$, the mode amplitudes of the spinning cylinders in this case can be expressed as [24]:

$$p_1 = |A_1| = \left| \frac{i\sqrt{\gamma_c} A_{in}}{\omega - (\omega_0 - \frac{i\Gamma}{2})} \right|, \quad p_2 = |A_2| = \left| \frac{i\sqrt{\gamma_c} A_{in}}{\omega - (\omega_0 - \frac{i\Gamma}{2})} + \frac{i\sqrt{\gamma_c} A_{in} \kappa_{21}}{[\omega - (\omega_0 - \frac{i\Gamma}{2})]^2} \right|, \quad (6)$$

where $\Gamma = \gamma + \gamma_c$ is the total loss. As seen in Eq. (6), the mode amplitude of cylinder 1 (i.e., $p_1$) has the same expression as the first term of the mode amplitude of cylinder 2 (i.e., $p_2$). The mode amplitude of cylinder 2 has an additional term with a second-order pole interfering with the first term, which can give rise to an asymmetry linear shape of the modal spectrum. The coupling parameter $\kappa_{21}$ does not appear in the expression of $p_1$, and it only affects $p_2$ via the second term. Since the coupling is achieved via the waveguide channel, we have $\kappa_{21} = -i\kappa_{21}^0 \exp(ik_{wg}d)$, where $\kappa_{21}^0$ is the positive real-valued coupling strength and $k_{wg}$ is the propagation constant of the waveguide mode. We note that the coupling of CCW mode of the spinning cylinders share similar physics except that it carries opposite spin that leads to $\kappa_{21} = 0$ and $\kappa_{12} \neq 0$. Importantly, the frequency splitting of the CW and CCW modes guarantees that spin flipping due to perturbations will not lead to the coupling of the two modes in contrast to the case of stationary resonators [24].

## 3. Results and discussions

*3.1 Exceptional point in two coupled spinning cylinders*

For the configuration shown in Fig. 1(a), we consider the two cylinders are made of silicon (relative permittivity 11.9) and the slab waveguide is made of silicon dioxide (relative permittivity 2.1). The cylinders have radii $R = 300$ nm and rotate in clockwise direction with angular velocity $\Omega$. The thickness of the slab waveguide is 400 nm. We conducted full-wave simulations of the structure by using COMSOL Multiphysics [35], assuming absorbing boundary conditions at both ends of the waveguide to suppress the reflections of the waveguide mode.

Figure 1(b) shows the mode amplitude as a function of the excitation frequency, which is defined as $A = \int |E| dS / (\pi R^2) = \overline{|E|}$. As shown by the blue dashed line, at zero spinning speed ($\Omega = 0$), a resonance appears at the frequency $f = 288.5$ THz, which corresponds to a pair of degenerated chiral modes (i.e., CW and CCW modes) with $m = \pm 4$. The $E_z$ field patterns of the chiral modes are shown in Fig. 1(c) and (d), respectively. At the normalized spinning speed $\Omega R / c = 0.01$, the two modes are spectrally separated due to the Sagnac effect [32], as shown by the red solid lines. In the following, we will focus on the EP associated with the CW mode of the two cylinders. The physics apply equally to the EP of the CCW mode.

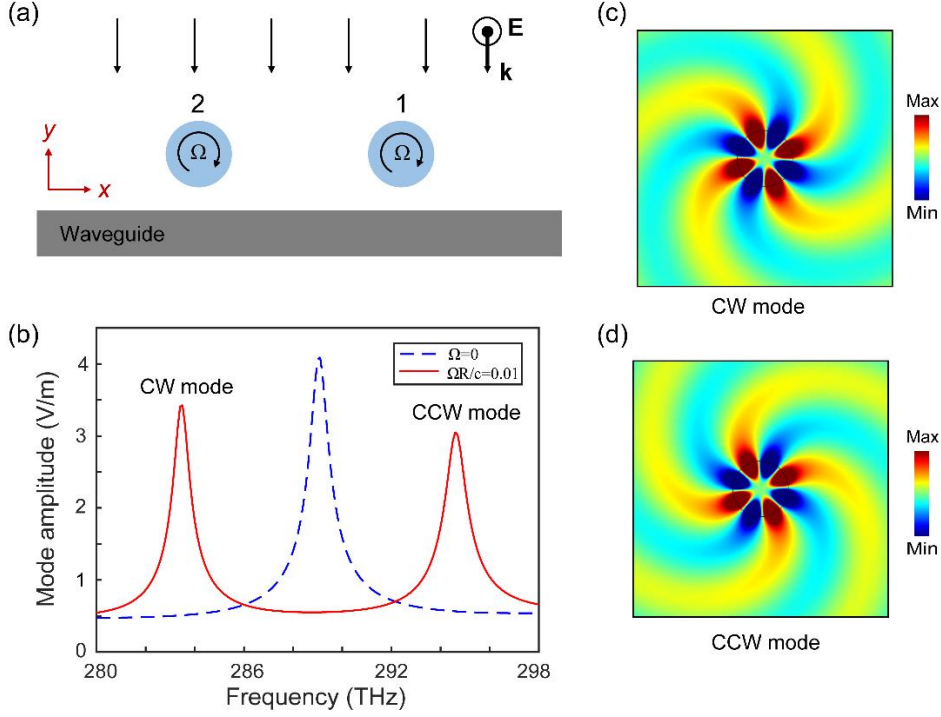

Fig. 1. (a) Two identical spinning cylinders couple via a waveguide under plane wave excitation. (b) Frequency splitting of the chiral modes in the cylinder for spinning speed $\Omega = 0$ and $\Omega R/c = 0.01$. (c), (d) $E_z$ field patterns for the CW mode and CCW mode, respectively.

We first simulated the mode amplitudes of the two cylinders ($p_1$ and $p_2$) for different distances $d$ at 4.8 μm, 4.9 μm, 4.95 μm, 5.0 μm, 5.05 μm, 5.1 μm, and 5.2 μm. The results are shown in Fig. 2(a) as the dot-symbol lines. The solid lines denote the analytical results obtained via fitting the numerical results to the CMT expressions in Eq. (6), which agree well with the simulation results. Evidently, $p_1$ exhibits a typical Lorentzian profile, while $p_2$ has an asymmetric profile that varies with $d$. This is a consequence of the unidirectional coupling induced by spin-orbit interaction. This unidirectionality can be verified by calculating the output power at both ends of the waveguide. As shown in Fig. 2(b), the output power at the right end vanishes, while the output power at the left end undergoes variation for different $d$ values due to the interference effect associated with $p_2$. At $\omega = \omega_0$, the expression for $p_2$ is reduced to $\left|4\sqrt{\gamma_c} A_{in} (\Gamma/2 - i\kappa_{21})/\Gamma^2\right|$, where $\kappa_{21} = -i\kappa_{21}^0 \exp(ik_{wg}d)$ depends on the value of $d$. Figure 2(c) shows the evolution of $Z(d) = \Gamma/2 - i\kappa_{21}$ in the complex plane as $d$ increases (the arrow denotes the evolution direction). The red dots correspond to the cases in Fig. 2(a) with $d$ increasing from $d = 4.8$ μm to $d = 5.2$ μm, which are obtained by the CMT fitting. When $d$ is a half-integer multiple of $\lambda_{wg}$, $Z(d)$ has a maximum value $\Gamma/2 + \kappa_{21}^0$, corresponding to constructive interference that gives rise to the higher peak of $p_2$ compared to $p_1$. When $d$ is an integer multiple of $\lambda_{wg}$, $Z(d)$ has a minimum value $\Gamma/2 - \kappa_{21}^0$, which corresponds to destructive interference that gives rise to a dip of $p_2$. Figure 2(d) and 2(e) show the electric field patterns at the constructive and destructive interferences, respectively (corresponding to the data points labelled as (d) and (e) in Fig. 2(a), respectively). As can be seen, the field magnitude in cylinder 1 is similar in both cases. In contrast, the field magnitude in cylinder 2 is enhanced in Fig. 2(d) and suppressed in Fig. 2(e). In addition, the coupling between the cylinders is obviously unidirectional, i.e., the guided wave propagates from cylinder 1 to cylinder 2.

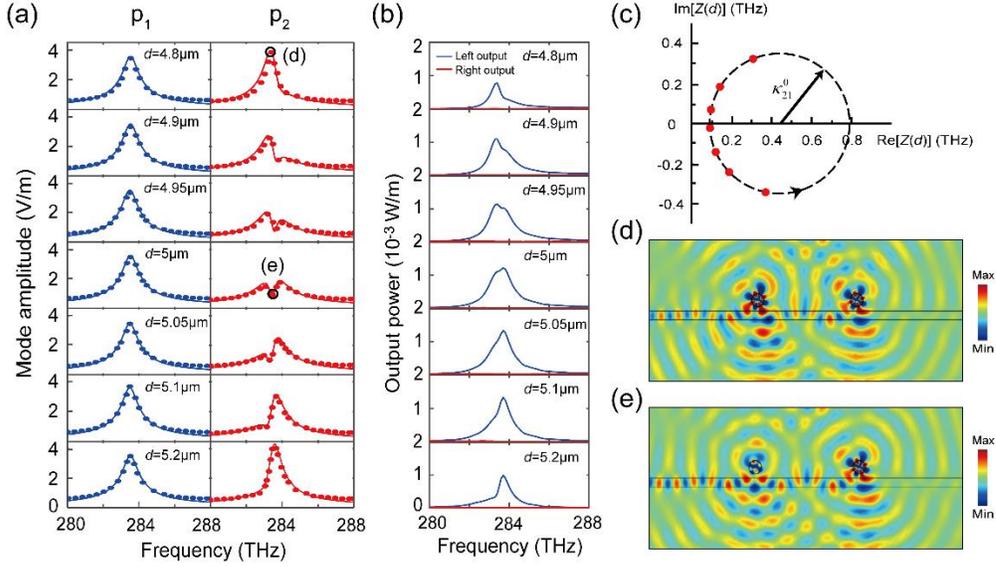

Fig. 2. (a) Mode amplitudes for cylinder 1 (left column) and cylinder 2 (right column) at $d$ = 4.8 μm, 4.9 μm, 4.95 μm, 5.0 μm, 5.05 μm, 5.1 μm, and 5.2 μm. Dots denote the simulation results and lines denote the fitting results of the CMT. (b) Output power at the left (blue line) and right (red line) ends of the waveguide. (c) Evolution of $Z(d) = \Gamma/2 - i\kappa_{21}$ on the complex plane as $d$ varies. The dashed circle denotes the evolution trajectory, and the arrow shows the evolution direction when $d$ increases. (d), (e) $E_z$ field patterns of the system corresponding to the data points labelled as (d) and (e) in panel (a), respectively.

*3.2 Realization of higher-order exceptional points*

Higher-order EPs can be straightforwardly realized in our system by placing more identical spinning cylinders on the waveguide. For $N$ cylinders coupled via the waveguide, the rate equations can be expressed as [24,33,34]:

$$\frac{da_i}{dt} = -i\omega_0 a_i - \frac{\Gamma}{2} a_i - i\kappa_{ij} \sum_{j=1}^{i-1} a_j - \sqrt{\gamma_c} a_{in}. \tag{7}$$

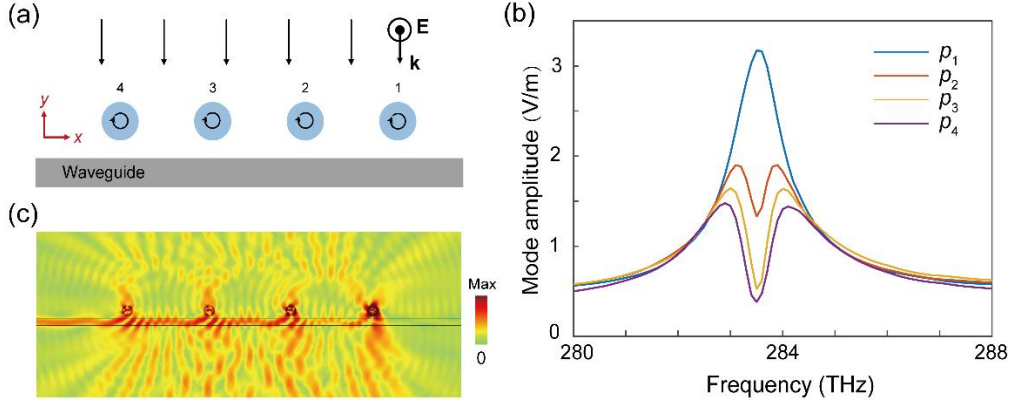

Fig. 3. (a) Unidirectional coupling of four spinning cylinders to achieve a higher-order EP. (b) Mode amplitudes for cylinder 1,2,3,4 with equal separation of $d = 4.99$ μm. (c) The electric filed amplitude at the resonant frequency.

The corresponding effective Hamiltonian is

$$H = \begin{bmatrix} \omega_0 - \dfrac{i}{2}\Gamma & 0 & \cdots & 0 \\ \kappa_{21} & \omega_0 - \dfrac{i}{2}\Gamma & \cdots & 0 \\ \vdots & \vdots & \ddots & \vdots \\ \kappa_{N1} & \kappa_{N2} & \cdots & \omega_0 - \dfrac{i}{2}\Gamma \end{bmatrix}, \tag{8}$$

where all the coupling parameters $\kappa_{ij}$ with $i < j$ vanish due to the unidirectional coupling. The above Hamiltonian has $N$ eigenvalues coalesced at $\omega = \omega_0 - i\Gamma/2$, corresponding to a $N^{\text{th}}$-order EP. To explore the properties of the system at this EP, we consider the case with $d = n\lambda_0$ and $n$ being an arbitrary nonzero integer. In this case, the coupling parameters reduce to $\kappa_{ij} = \kappa_{21} = -i\kappa_{21}^0$, where $\kappa_{21}^0$ is the coupling strength defined in Section 3.1. The mode amplitude of the spinning cylinders can be expressed as [24]:

$$p_i = |A_i| = \left| \frac{2\sqrt{\gamma_c} A_{in}}{\Gamma} \left( \frac{\Gamma - 2\kappa_{21}^0}{\Gamma} \right)^{i-1} \right| \tag{9}$$

Because $\Gamma$ and $\kappa_{21}^0$ both take positive reals and $\Gamma > 2\kappa_{21}^0$, it is clear that $\left|(\Gamma - 2\kappa_{21}^0)/\Gamma\right| < 1$. Consequently, we obtain $p_i < p_{i-1}$, which indicates that the mode amplitudes of the cylinders decrease from the right to the left.

To illustrate the high-order EP effect, we consider four spinning cylinders that are uniformly separated with a distance $d = 4.99$ μm, as shown in Fig. 3(a). The same linearly polarized plane wave is employed to excite the system. We numerically calculated the mode amplitudes of the cylinders, and the results are shown in Fig. 3(b). It is seen that only cylinder 1 exhibits a Lorentzian profile while the others are suppressed due to destructive interferences. Importantly, at the resonance frequency of the EP, the mode amplitudes monotonically decrease from cylinder 1 to cylinder 4, as predicted by Eq. (9). Figure 3(c) shows the electric field amplitude at the resonant frequency. We see that the field amplitude in the cylinders does reduce from right to left, and the guided wave propagates unidirectionally to the left side. This phenomenon may find applications in directional energy harvesting and spin-dependent optical switches.

*3.3 Effect of the exceptional points on optical isolation*

A major feature of the considered system is the co-existence of non-reciprocity and non-Hermicity and thus their interplay in determining the light scattering by the cylinders and light propagation in the waveguide. To explore this feature, we consider the system under the port excitation (TE mode) at the right end of the waveguide, as shown in Fig. 4(a), and the system is working at the frequency of the CW mode. We use COMSOL to simulate the mode amplitudes of cylinder 1 and cylinder 2 while varying the distance $d$ from 5.3 μm to 5.8 μm, and the results are shown in Fig. 4(b) and 4(c), respectively. As expected, cylinder 1 exhibits the symmetric Lorentzian response at all $d$ values, which is similar to the results of plane wave excitation in Figs. 2 and 3. Interestingly, the mode amplitude of cylinder 2 exhibits destructive interference for all values of $d$, in contrast to the results in Fig. 2(a) under plane wave excitation. This phenomenon is attributed to the phase difference of the incident wave at the two cylinders. The mode amplitude of cylinder 2 in this case can be expressed as:

$$p_2 = \left| \frac{i\sqrt{\gamma_c} A_{in2}}{\omega - (\omega_0 - \frac{i\Gamma}{2})} + \frac{i\sqrt{\gamma_c} A_{in1} \kappa_{21}}{[\omega - (\omega_0 - \frac{i\Gamma}{2})]^2} \right|, \quad (10)$$

where $A_{in1}$, $A_{in2}$ denote the complex amplitudes of the incident wave at the position of the two cylinders. For a separation of $d$ between the two cylinders, we have $A_{in1} = A_{in2} \exp(-ik_{wg}d)$ and $\kappa_{21} = -i\kappa_{21}^0 \exp(ik_{wg}d)$. Therefore, the product $A_{in1}\kappa_{21}$ is independent of the distance $d$, and the interference of the two terms in $p_2$ is always destructive.

To understand the effect of non-Hermiticity and non-reciprocity on light propagation in the waveguide, we numerically compute the transmission of the waveguide under excitation at the left and right inputs. The results are shown in Fig. 4(d) for different numbers of cylinders $N = 1, 2, 3, 4$. The case of $N = 1$ corresponds to conventional optical isolation realized by single spinning resonator [30]. The other cases with $N = 2, 3, 4$ correspond to EPs of order $N$. As can be seen, approximately total transmission happens for the cases with input from left (dashed lines) while the transmission shows a dip for the cases with input from right (solid lines). We define the optical isolation ratio as the transmission contrast between "input from left" and "input from right". The results are shown in Fig. 4(e). We see that the isolation ratio increases with the increase of the cylinder number $N$ (i.e., the order of the EPs). In conventional optical isolation systems, increasing the number of nonreciprocal components could also increase the isolation ratio. However, the couplings between multiple components can lead to complex interferences that render the isolation performance sensitive to the variation of system parameters. In our proposed system, the enhancement of isolation ratio at higher-order EPs is guaranteed by the unidirectional coupling under spin-orbit interaction, thus the destructive interference is robust against variation of coupling distance.

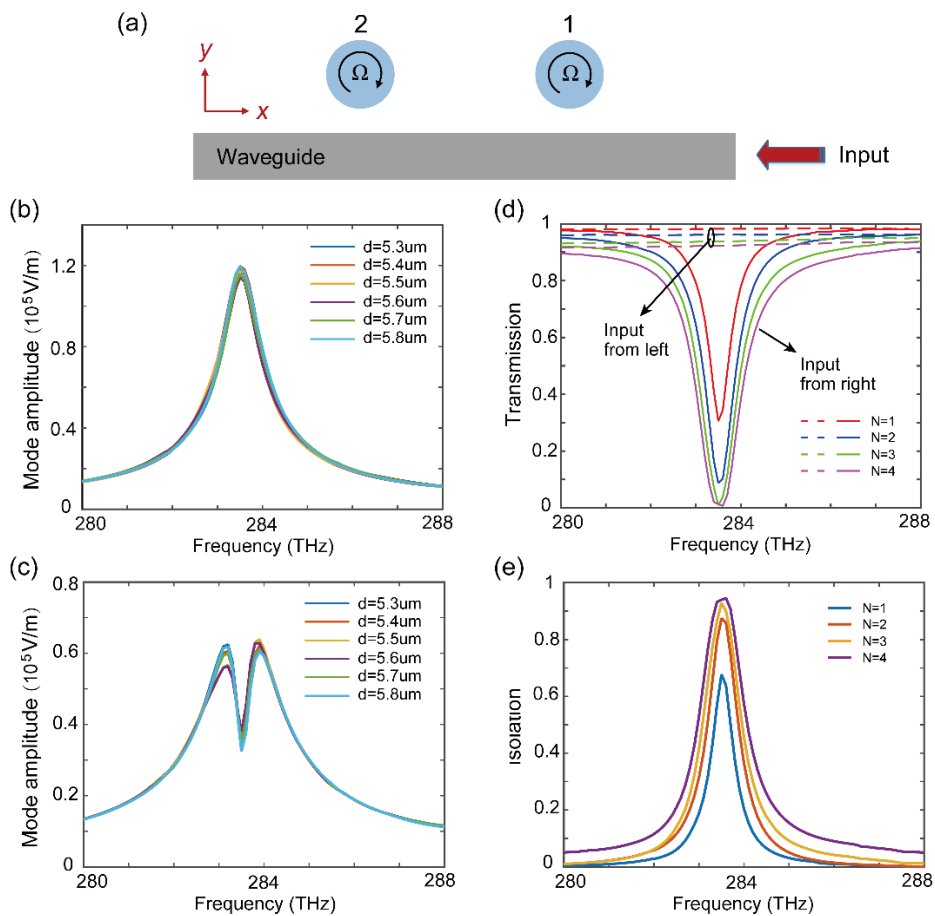

Fig. 4. (a) Optical isolation induced by coupled spinning cylinders under the port excitation. (b), (c) Mode amplitudes of the cylinder 1 and cylinder 2, respectively, for different distance $d$. (d) Transmission under left and right port excitations. (e) Isolation ratio for different number of cylinders.

## 4. Conclusions

To summarize, we showed that EPs of arbitrary order can be realized in spinning cylinders coupled through a dielectric waveguide. In contrast to previous studies employing spin-orbit interactions to achieve EPs, the proposed configuration does not rely on selective excitation of the chiral modes and thus is robust against perturbations that can flip the spin. This is attributed to the frequency splitting between a pair of chiral modes induced by spinning of the cylinders. By using full-wave numerical simulations and CMT, we demonstrated the asymmetric response of the cylinders at the EPs and the enhanced optical isolation of the waveguide mode. The study can be extended to higher-order chiral modes (i.e., whispering gallery modes) in microcavities, where the spinning speed can be significantly lowered and experimental implementations are feasible [36–39]. Our study contributes to the understanding of light-matter interactions in moving medium and may find applications in optical isolations, topological photonics, and chiral quantum optics.

**Funding.** Research Grants Council of the Hong Kong Special Administrative Region, China (CityU 11301820, CityU 21302018, C6013-18G); National Natural Science Foundation of China (NSFC) (11904306).

## References

1. L. Feng, R. El-Ganainy, and L. Ge, "Non-Hermitian photonics based on parity–time symmetry," Nat. Photonics **11**, 752–762 (2017).
2. W. D. Heiss, "Exceptional points of non-Hermitian operators," J. Phys. A. Math. Gen. **37**, 2455–2464 (2004).
3. W. D. Heiss and A. L. Sannino, "Avoided level crossing and exceptional points," J. Phys. A. Math. Gen. **23**, 1167–1178 (1990).
4. R. El-Ganainy, K. G. Makris, M. Khajavikhan, Z. H. Musslimani, S. Rotter, and D. N. Christodoulides, "Non-Hermitian physics and PT symmetry," Nat. Phys. **14**, 11–19 (2018).
5. B. Peng, Ş. K. Özdemir, S. Rotter, H. Yilmaz, M. Liertzer, F. Monifi, C. M. Bender, F. Nori, and L. Yang, "Loss-induced suppression and revival of lasing," Science **346**, 328–332 (2014).
6. L. Chang, X. Jiang, S. Hua, C. Yang, J. Wen, L. Jiang, G. Li, G. Wang, and M. Xiao, "Parity–time symmetry and variable optical isolation in active–passive-coupled microresonators," Nat. Photonics **8**, 524–529 (2014).
7. H. Zhang, R. Huang, S.-D. Zhang, Y. Li, C.-W. Qiu, F. Nori, and H. Jing, "Breaking Anti-PT Symmetry by Spinning a Resonator," Nano Lett. **20**, 7594–7599 (2020).
8. A. Lupu, H. Benisty, and A. Degiron, "Using optical PT-symmetry for switching applications," Photonics Nanostructures - Fundam. Appl. **12**, 305–311 (2014).
9. X. Zhang, X. Wang, and C. T. Chan, "Switching Terahertz Waves using Exceptional Points," Phys. Rev. Appl. **10**, 1 (2018).
10. A. Lupu, H. Benisty, and A. Degiron, "Switching using PT symmetry in plasmonic systems: positive role of the losses," Opt. Express **21**, 21651–21668 (2013).
11. X.-L. Zhang, S. Wang, W.-J. Chen, and C. T. Chan, "Exceptional points and symmetry recovery in a two-state system," Phys. Rev. A **96**, 022112 (2017).

12. X.-L. Zhang, S. Wang, B. Hou, and C. T. Chan, "Dynamically Encircling Exceptional Points: In situ Control of Encircling Loops and the Role of the Starting Point," Phys. Rev. X **8**, 021066 (2018).
13. P. Miao, Z. Zhang, J. Sun, W. Walasik, S. Longhi, N. M. Litchinitser, and L. Feng, "Orbital angular momentum microlaser," Science **353**, 464–467 (2016).
14. M. Brandstetter, M. Liertzer, C. Deutsch, P. Klang, J. Schöberl, H. E. Türeci, G. Strasser, K. Unterrainer, and S. Rotter, "Reversing the pump dependence of a laser at an exceptional point," Nat. Commun. **5**, 4034 (2014).
15. Z. Lin, A. Pick, M. Lon, and A. W. Rodriguez, "Enhanced Spontaneous Emission at Third-Order Dirac Exceptional Points in Inverse-Designed Photonic Crystals," Phys. Rev. Lett. **117**, 107402 (2016).
16. W. D. Heiss, "Chirality of wavefunctions for three coalescing levels," J. Phys. A Math. Theor. **41**, 244010 (2008).
17. K. Ding, G. Ma, M. Xiao, Z. Q. Zhang, and C. T. Chan, "Emergence, Coalescence, and Topological Properties of Multiple Exceptional Points and Their Experimental Realization," Phys. Rev. X **6**, 021007 (2016).
18. T. Gao, E. Estrecho, K. Y. Bliokh, T. C. H. Liew, M. D. Fraser, S. Brodbeck, M. Kamp, C. Schneider, S. Höfling, Y. Yamamoto, F. Nori, Y. S. Kivshar, A. G. Truscott, R. G. Dall, and E. A. Ostrovskaya, "Observation of non-Hermitian degeneracies in a chaotic exciton-polariton billiard," Nature **526**, 554–558 (2015).
19. H. Hodaei, A. U. Hassan, S. Wittek, H. Garcia-Gracia, R. El-Ganainy, D. N. Christodoulides, and M. Khajavikhan, "Enhanced sensitivity at higher-order exceptional points," Nature **548**, 187–191 (2017).
20. W. Chen, Ş. Kaya Özdemir, G. Zhao, J. Wiersig, and L. Yang, "Exceptional points enhance sensing in an optical microcavity," Nature **548**, 192–196 (2017).
21. L. Feng, R. El-Ganainy, and L. Ge, "Non-Hermitian photonics based on parity-time symmetry," Nat. Photonics **11**, 752–762 (2017).
22. R. El-ganainy, K. G. Makris, M. Khajavikhan, Z. H. Musslimani, S. Rotter, and D. N. Christodoulides, "Non-Hermitian physics and PT symmetry," Nat. Phys. **14**, 11–19 (2018).
23. B. Peng, Ş. K. Özdemir, M. Liertzer, W. Chen, J. Kramer, H. Yılmaz, J. Wiersig, S. Rotter, and L. Yang, "Chiral modes and directional lasing at exceptional points," Proc. Natl. Acad. Sci. **113**, 6845–6850 (2016).
24. S. Wang, B. Hou, W. Lu, Y. Chen, Z. Q. Zhang, and C. T. Chan, "Arbitrary order exceptional point induced by photonic spin–orbit interaction in coupled resonators," Nat. Commun. **10**, 832 (2019).
25. J. Wiersig, "Enhancing the Sensitivity of Frequency and Energy Splitting Detection by Using Exceptional Points: Application to Microcavity Sensors for Single-Particle Detection," Phys. Rev. Lett. **112**, 203901 (2014).
26. K. Y. Bliokh, F. J. Rodríguez-Fortuño, F. Nori, and A. V Zayats, "Spin–orbit interactions of light," Nat. Photonics **9**, 796–808 (2015).

27. S. B. Wang and C. T. Chan, "Lateral optical force on chiral particles near a surface," Nat. Commun. **5**, 3307 (2014).
28. K. Y. Bliokh, D. Smirnova, and F. Nori, "Quantum spin Hall effect of light," Science **348**, 1448–1451 (2015).
29. K. Y. Bliokh, A. Y. Bekshaev, and F. Nori, "Extraordinary momentum and spin in evanescent waves," Nat. Commun. **5**, 3300 (2014).
30. H. Shi, Y. Cheng, Z. Yang, Y. Chen, and S. Wang, "Optical isolation induced by subwavelength spinning particle via spin-orbit interaction," Phys. Rev. B **103**, 094105 (2021).
31. H. Minkowski, "Die Grundgleichungen für die elektromagnetischen Vorgänge in bewegten Körpern," Nachrichten von der Gesellschaft der Wissenschaften zu Göttingen, Math. Klasse **1908**, 53–111 (1908).
32. H. Shi, Z. Xiong, W. Chen, J. Xu, S. Wang, and Y. Chen, "Gauge-field description of Sagnac frequency shift and mode hybridization in a rotating cavity," Opt. Express **27**, 28114–28122 (2019).
33. S. Fan, W. Suh, and J. D. Joannopoulos, "Temporal coupled-mode theory for the Fano resonance in optical resonators," J. Opt. Soc. Am. A **20**, 569–572 (2003).
34. Wonjoo Suh, Zheng Wang, and Shanhui Fan, "Temporal coupled-mode theory and the presence of non-orthogonal modes in lossless multimode cavities," IEEE J. Quantum Electron. **40**, 1511–1518 (2004).
35. www.comsol.com.
36. J. Gieseler, B. Deutsch, R. Quidant, and L. Novotny, "Subkelvin Parametric Feedback Cooling of a Laser-Trapped Nanoparticle," Phys. Rev. Lett. **109**, 103603 (2012).
37. J. Ahn, Z. Xu, J. Bang, Y.-H. Deng, T. M. Hoang, Q. Han, R.-M. Ma, and T. Li, "Optically Levitated Nanodumbbell Torsion Balance and GHz Nanomechanical Rotor," Phys. Rev. Lett. **121**, 033603 (2018).
38. J. Ahn, Z. Xu, J. Bang, P. Ju, X. Gao, and T. Li, "Ultrasensitive torque detection with an optically levitated nanorotor," Nat. Nanotechnol. **15**, 89–93 (2020).
39. S. Maayani, R. Dahan, Y. Kligerman, E. Moses, A. U. Hassan, H. Jing, F. Nori, D. N. Christodoulides, and T. Carmon, "Flying couplers above spinning resonators generate irreversible refraction," Nature **558**, 569–572 (2018).